\documentclass[preprint]{aastex}
\usepackage{graphicx}
\usepackage{natbib,times}
\usepackage{lscape}

\def\Msolar{M_{\odot}}

\shorttitle{}
\shortauthors{}
\begin{document}

\title{Constraining the coalescence rate of supermassive black-hole binaries\\
using pulsar timing}

\author{Z. L. Wen\altaffilmark{1}, F. A. Jenet\altaffilmark{2}, D. Yardley\altaffilmark{3,4}, 
G. B. Hobbs\altaffilmark{4}, R. N. Manchester\altaffilmark{4}}

\altaffiltext{1}{National Astronomical Observatories, Chinese Academy
of Sciences, Jia-20 DaTun Road, ChaoYang District, Beijing 100012,
China (zhonglue@bao.ac.cn)} \altaffiltext{2}{Center for Gravitational Wave Astronomy,
University of Texas at Brownsville, TX 78520 (merlyn@phys.utb.edu)}
\altaffiltext{3}{Sydney Institute for Astronomy (SIfA), School of
  Physics, The University of Sydney, NSW 2006, Australia}
\altaffiltext{4}{CSIRO Astronomy and Space Science, Australia
  Telescope National Facility, PO Box 77, Epping NSW 1710, Australia} 


\begin{abstract}

  Pulsar timing observations are used to place constraints on the rate
  of coalescence of supermassive black-hole (SMBH) binaries as a
  function of mass and redshift. In contrast to the indirect
  constraints obtained from other techniques, pulsar timing
  observations provide a direct constraint on the black-hole merger
  rate. This is possible since pulsar timing is sensitive to the
  gravitational waves (GWs) emitted by these sources in the final
  stages of their evolution. We find that upper bounds calculated from
  the recently published Parkes Pulsar Timing Array data are just above
  theoretical predictions for redshifts below 10. In the future, with
  improved timing precision and longer data spans, we show that a
  non-detection of GWs will rule out some of the available parameter
  space in a particular class of SMBH binary merger models. We also
  show that if we can time a set of pulsars to 10\,ns timing accuracy,
  for example, using the proposed Square Kilometre Array, it should be
  possible to detect one or more individual SMBH binary systems.

\end{abstract}


\section{Introduction}

Pulsar timing observations \citep[for a review of the techniques,
  see][]{lk05,ehm06} provide a unique opportunity to study
low-frequency ($10^{-9}$--$10^{-7}$ Hz) gravitational waves
\citep[GWs;][]{saz78,det79,bcr83,hd83,fb90,ktr94,jhl+05}. Previous
work \citep{rt83,ktr94,lom02,jhv+06} placed upper limits on a
stochastic background of GWs. These limits were reported in terms of
either the amplitude of the GW characteristic strain spectrum,
$h_c(f)$, or the normalized GW energy density, $\Omega_{\rm gw}(f)$.

In recent years, researchers have proposed that supermassive
black-hole (SMBH) binary systems distributed throughout the universe
will be a source of GWs detectable using pulsar timing techniques
\citep[e.g.,][]{jb03,wl03,eins04,shm+04,svc08}. The detection, or
non-detection, of such GWs provides a constraint on the rate of
coalescence of SMBH binary systems. We emphasize that such constraints
are model-independent as opposed to the indirect constraints that can
be inferred from observed galaxy distributions.

We will show that existing pulsar data sets do not provide stringent
constraints on the coalescence rate.  However, future data sets from
the Parkes Pulsar Timing Array \citep[PPTA;][]{man08,hbb+09} and
similar North American \citep[NANOGrav;][]{jen09} and European timing
array \citep[EPTA;][]{skl+06} projects aim to produce data sets on 20 or
more pulsars with root-mean-square (rms) timing residuals close to
100\,ns.  In the longer term, we expect that pulsar timing array projects using
future telescopes such as the Square Kilometre Array
(SKA)\footnote{See www.skatelescope.org} will be able to time many
hundreds of pulsars with exquisite timing precision.

The outline of this paper is as follows. In \S~2, the physics of GW
emission from SMBH binary systems is reviewed together with the
effects of GWs on pulsar timing. We describe how to constrain the
coalescence rate using two different techniques, one valid when there
is a large number of expected sources and the other valid when only a
few sources are expected. In \S~3 we show the recent and projected
rate constraints for various different observing systems. These
observationally constrained rates are then compared to the rates
implied by local galaxy-merger observations. This work is summarized
in section \S~4.

\section{Constraining the SMBH Merger Rate}

We define a SMBH as a black hole with mass greater than
$10^6~\Msolar$. There is abundant evidence that such SMBHs exist both
nearby \citep[e.g.,][]{kr92,mmh+95} and at high redshifts $z \sim$ 6
\citep[e.g.,][]{fnl+01}. Two orbiting SMBHs resulting from a galaxy
merger would emit large amplitude
GWs. Such GW sources are important targets for space-based detectors
such as the Laser Interferometer Space Antenna (LISA) and pulsar
timing arrays \citep[e.g.,][]{wl03,eins04,shm+04}.  Observational GW
astronomy can be used to resolve whether SMBH binary systems form and
whether the two black holes can get close enough to emit detectable
GWs.

\begin{figure}
\epsscale{0.5}
\plotone{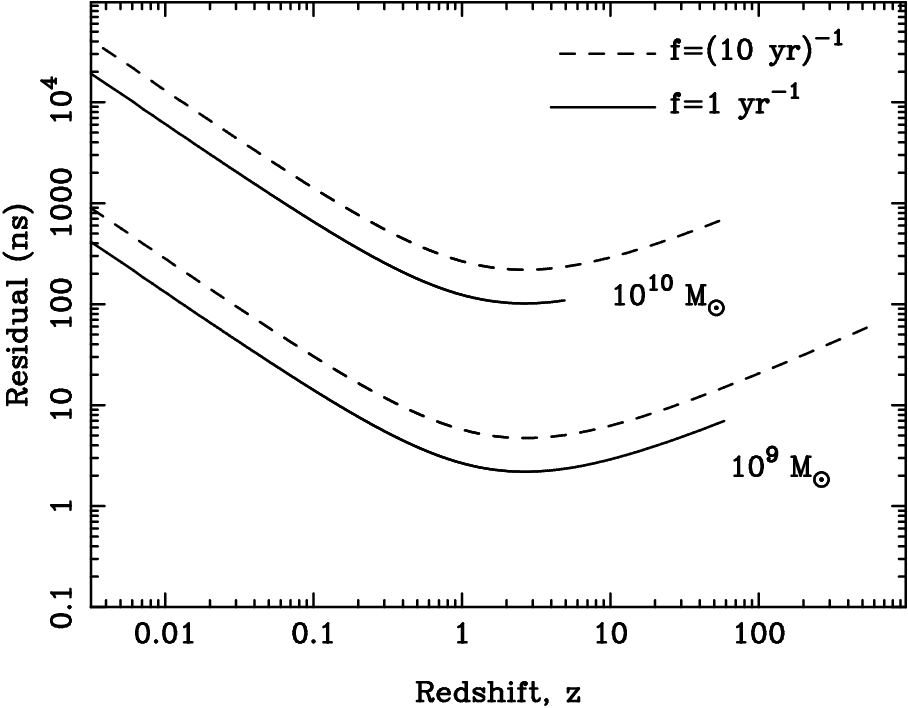}
\caption{\label{rvsz}Induced timing residual amplitudes versus 
redshift of a system with a given observed frequency and chirp mass.}
\vspace{2.em}
\end{figure}

This work focuses on the rate constraints measurable by pulsar timing
observations. In order to determine the coalescence rate, one needs to
know the expected GW amplitude emitted by a SMBH binary. This is given
by \citep{tho87}
\begin{equation}
h_s=4\sqrt{\frac{2}{5}}\frac{(GM_c)^{5/3}}{c^4D(z)}[\pi f(1+z)]^{2/3},
\label{amp}
\end{equation}
where $M_c$ is the ``chirp mass'' of the SMBH binary given by $M_c
=(M_1M_2)^{3/5}(M_1+M_2)^{-1/5}$, $M_1$ and $M_2$ are the individual
black hole masses, $f$ is observed GW frequency, $D(z)$ is the
comoving distance to the system
\begin{equation}
D(z)=\frac{c}{H_0}\int_0^z
\frac{dz'}{E(z')},
\label{distance}
\end{equation}
with $E(z)=\sqrt{\Omega_{\Lambda}+\Omega_m(1+z)^3}$ for a $\Lambda$CDM
cosmological model (hereafter we adopt $H_0=72~{\rm km~s^{-1}~Mpc^{-1}}$,
$\Omega_m=0.3$, $\Omega_{\Lambda}=0.7$). GWs from such a source will induce 
sinusoidally oscillating arrival-time variations whose amplitude,
$\Delta t$, is given by
\citep[][and references therein]{jll+04}:
\begin{equation}
\Delta t = \frac{h_s}{\omega} [1+\cos(\theta)]\sin(2\phi)\sin\{\omega D[1 - \cos(\theta)]/2c\},
\end{equation}
where $\omega$ is the GW frequency in radians/s, $\theta$ is the angle
on the sky between the pulsar direction and the GW source direction,
$\phi$ is the GW polarization angle, and $D$ is the distance to the
pulsar. The maximum induced timing residual amplitude, $h_s/\omega$,
is plotted in Figure \ref{rvsz} as a function of redshift for chirp
masses of $10^9$ and $10^{10}~\Msolar$ and observed GW frequencies of
(1 year)$^{-1}$ and (10 year)$^{-1}$. The most notable feature in this
Figure is that these amplitude curves are not monotonically decreasing
with increasing redshift. The reason for this is that the observing
frequency is held fixed and the frequency in the frame of the
emitting system increases with increasing redshift. For a binary
system, the emitted GW amplitude increases with increasing
frequency. For large $z$, this increase is faster than the decrease
due to increasing comoving distance $D(z)$. Also note that the curves
cutoff at large $z$. This is because there is a maximum orbital
frequency allowed before the black holes plunge together. This maximum
frequency was taken to be $c^3/(6^{3/2}\pi GM)$ assuming a circular
orbit \citep{hug02}. Here, $M=M_1+M_2$ is the total mass of a binary
system.

Rms timing residuals (for a typical 1-hour observation) for the best
pulsar data sets are currently around 100\,ns. Multiple observations
combined with improved systems will bring the effective sensitivity down
to around 10\,ns. In this case, Figure \ref{rvsz} shows that pulsar
timing will be sensitive to individual SMBH binary systems with chirp
masses greater than about $10^9~\Msolar$. Figure \ref{rvsz} also shows
that this sensitivity extends to large redshifts. This fact greatly
increases the chances of detecting individual sources. An ensemble of
lower-mass SMBH binary systems will be detectable as a stochastic
background if there is a large enough population of sources.

Since pulsar timing techniques are sensitive to SMBH binary systems up
to high redshifts, the non-detection of any sources, either individual
binaries or a background generated by an ensemble of binary systems,
may be used to place a direct constraint on $d^2R/dM_c dz$, the
sky-averaged rate of coalescence of binary SMBHs per unit chirp mass,
$M_c$, per unit redshift $z$. The total number of binary SMBHs with
chirp mass between $M_c$ and $M_c + \Delta M_c$ and located between
$z$ and $z+\Delta z$ merging between time $t$ and $t + \Delta t$ is
given by $\Delta M_c \Delta z \Delta t d^2R/dM_cdz$. Constraints
placed on this quantity may be used to rule out various binary SMBH
formation models.

We present two methods for determining the coalescence rate from
pulsar timing data. The first is valid when the rate is high enough
that the GWs form a stochastic background (the ``stochastic
constraint'') and there is a large number of sources per resolvable
frequency bin.  The second method (the ``Poissonian constraint'')
provides an estimate of the coalescence rate when the stochastic
constraint does not hold.  For a real data set it is practical first
to assume the stochastic constraint, determine the coalescence rate
and check whether the rate is high enough for the assumption to be
valid.  If not then the Poissonian constraint should be used.

\subsection{Stochastic constraint}

Here, it is assumed that a large number of SMBH binary sources form an
incoherent background of GWs. The power spectrum of such a background
is given by \citet{jb03}
\begin{equation}
P(f) = \int_{0}^{\infty} \int_{0}^{\infty} h_s(f,M_c,z)^2 \frac{d^2R}{dz dM_c} \left(\frac{df}{dt}\right)^{-1} dz dM_c,
\label{power1}
\end{equation}
where $h_s(f,M_c,z)$ is given by Equation~\ref{amp} and $df/dt$ is the
rate of change of the observed GW frequency. For the case of a binary
system evolving under general relativity alone, this is given by \citep{pm63}
\begin{equation}
\frac{df}{dt} = \frac{96}{5}\left(\frac{G M_c}{c^3}\right)^{5/3}\left(\pi f\right)^{8/3}f(1+z)^{5/3}.
\label{dfdt}
\end{equation}
Typically, a stochastic background of GWs is described by its
characteristic strain spectrum, $h_c(f)$, which is assumed to take on
a power law form:
\begin{equation}\label{eqn:bkgrd}
h_c(f)=A\Big(\frac{f}{f_{\rm yr}}\Big)^{\alpha},
\end{equation}
where $f_{\rm yr} = 1/(1 {\rm year})$ and $A$ is the characteristic
strain at a period of one year. For a background generated by SMBH
binaries, $\alpha=-2/3$ in frequency range
$10^{-9}$--$10^{-7}$ Hz \citep{jb03,wl03,shm+04,eins04}. Note that the
characteristic strain spectrum is related to the power spectrum by
$P(f) = h_c(f)^2/f$.

Pulsar timing data sets provide an upper bound, $A_{\rm up}$, on
$A$. Such bounds limit the power spectrum of the GW strain. The upper
bound, $P_{\rm up}(f)$, may be written as
\begin{equation}
P_{\rm up}(f) = \frac{A^2_{\rm up}}{f_{\rm yr}}\Big(\frac{f}{f_{\rm yr}}\Big)^{-7/3}.
\label{Pup}
\end{equation}
Since this is an upper bound, we have 
\begin{equation}
P(f) < P_{\rm up}(f).
\label{ub}
\end{equation}
In order to use Equations~\ref{power1} and \ref{ub} to obtain a
constraint on the differential rate of coalescence itself, the
integrand is rewritten in an equivalent form:
\begin{equation}
P(f) = \int_0^{\infty} \int_{-\infty}^{\infty} h_s(f,M_c,z)^2 \frac{d^2R}{d\lg(1+z) d\lg(M_c)} \left(\frac{df}{dt}\right)^{-1} d\lg(1+z) d\lg(M_c).
\label{power2}
\end{equation}
Note that both $d^2R/d\lg(1+z) d\lg(M_c)$ and $df/dt$ depend on $z$ and
$M_c$, although the explicit dependence is not written. A constraint
on $P(f)$ is a direct constraint on the integral in the above
expression. In order to obtain an estimate of the upper bound on the
integrand, we follow the same line of reasoning used to place
constraints on the differential energy density of gravitational waves
using bounds from big bang nucleosynthesis \citep{mag00}. First, we
note that the limits in the integral of equation \ref{power2} are from
0 to $\infty$ for $\lg(1+z)$ and from $-\infty$ to $\infty$ for
$\lg(M_c)$. Consider this integral over a small region bounded
by $\lg(M_{c_1})$ to $\lg(M_{c_2})$ and $\lg(1+z_1)$ to $\lg(1+z_2)$.
Denote this integral as $P_s(f)$ where
\begin{equation}
P_s(f) = \int_{\lg(M_{c_1})}^{\lg(M_{c_2})} \int_{\lg(1+z_1)}^{\lg(1+z_2)} h_s(f,M_c,z)^2 
\frac{d^2R}{d\lg(1+z) d\lg(M_c)} \left(\frac{df}{dt}\right)^{-1} d\lg(1+z) d\lg(M_c).
\end{equation}
The mean value theorem tells us that there exist values of $M_c$ and
$z$, written as $M_c^*$ and $z^*$, such that
\begin{equation}
P_s(f) = h_s(f,M_c^*,z^*)^2 \frac{d^2R}{d\lg(1+z) d\lg(M_c)}
\left(\frac{df}{dt}\right)^{-1} \Delta\lg(1+z) \Delta\lg(M_c),
\end{equation}
where $\Delta\lg(1+z) = \lg(1+z_2) - \lg(1+z_1)$ and $\Delta\lg(M_c) =
\lg(M_{c_2}) - \lg(M_{c_1})$. Next, we assume that the integrand
varies slowly over the region of integration so that $P_s(f)$ does
not change much as long as $M_c^*$ and $z^*$ are chosen within this
region. From this, we see that $P_s(f)$ is
approximately given by
\begin{equation}
P_s(f) \approx h_s(f,M_c,z)^2 \frac{d^2R}{d\lg(1+z) d\lg(M_c)}
\left(\frac{df}{dt}\right)^{-1} \Delta\lg(1+z)\Delta\lg(M_c)
\label{power_approx}
\end{equation}
for any value of $M_c \in [M_{c_1}, M_{c_2}]$ and $z \in
[z_1,z_2]$. Next, since the integrand in equation \ref{power2} is
positive definite, it follows that
\begin{equation}
P_s(f) \leq P(f).
\label{ub2}
\end{equation}
From equations \ref{Pup}, \ref{ub}, \ref{power_approx}, and \ref{ub2} we have:
\begin{equation}
h_s(f,M_c,z)^2 \frac{d^2R}{d\lg(1+z) d\lg(M_c)} \left(\frac{df}{dt}\right)^{-1} \Delta\lg(1+z)\Delta\lg(M_c)
\lesssim \frac{A^2_{\rm up}}{f_{\rm yr}}\Big(\frac{f}{f_{\rm yr}}\Big)^{-7/3},
\label{integrand_c}
\end{equation}
where we are free to choose any value for $M_c$ and $z$ provided that
the integrand is slowly varying over the appropriate region. Using the
known expressions for $h_s(f,M_c,z)$ and $df/dt$, one can use equation
\ref{integrand_c} to obtain a constraint on the
differential rate of coalescence. Assuming $\Delta\lg(M_c)=1$ and
$\Delta \lg(1+z) = 0.2$, the constraint takes the form
\begin{equation}
\frac{d^2 R}{d\lg(1+z) d\lg(M_c)} < 15 A^2_{\rm up} \frac{c^3 D(z)^2 (1 + z)^{1/3}}{(G M_c)^{5/3}}
(\pi f_{\rm yr})^{4/3},
\label{const_st}
\end{equation}
which is independent of frequency.  Therefore, a measured bound,
$A_{\rm up}$, can be used directly to constrain the SMBH coalescence
rate. Note that, if one believes that the integrand changes more rapidly with $z$
and/or $M_c$, one can choose a sufficiently small integration range
over which this assumption is true and then rescale the above
constraint.

This constraint is only valid when there are a large number of sources
emitting into the same frequency band at the same time. In order for
this to be true, the GW amplitude of each source must be much less
than the minimum detectable amplitude as determined by the statistical
properties of the pulsar timing data, otherwise a detection would
have been made. This reasoning leads to the following constraint:
$h_s(f,M_c,z)^2 \ll P_{\rm up}(f) \Delta f$, where $\Delta f$ is the
resolution bandwidth which is taken to be $1/T_{\rm obs}$. For
the purposes of making numerical estimates, the following constraint is
used
\begin{equation}
h_s(f,M_c,z)^2 \leq 0.1 \frac{P_{\rm up}(f)}{T_{\rm obs}}.
\end{equation}
For a fixed chirp mass and frequency, this expression is
a constraint on $z$. The most stringent constraint occurs when
$f=1/T_{\rm obs}$, the lowest observable frequency. Combining the above
with Equations~\ref{amp} and \ref{Pup} yields the following constraint
on the redshift
\begin{equation}
\frac{(1+z)^{2/3}}{D(z)} \leq 0.19\Big(\frac{A_{\rm up}}{10^{-14}}\Big)
\Big(\frac{T_{\rm obs}}{\rm yr}\Big)^{4/3}
\Big(\frac{M_c}{10^8~\Msolar}\Big)^{-5/3}{\rm Mpc}^{-1}.
\label{condition}
\end{equation}
The factor $(1+z)^{2/3}/D(z)$ decreases with $z$ until $z=$2.65,
where it starts to increase. Hence, there is a bounded redshift
interval over which the stochastic constraint is valid. For systems
outside of this range, the stochastic rate limit is not valid and the
Poisson rate limit discussed in the next section must be employed.

\subsection{Poisson constraint}

For this case, the sources are not numerous enough to form a
stochastic background, so they must be treated as individual
events. Assuming Poissonian statistics for the probability of an event
occurring, the probability that no events are detected is given by
$e^{-\langle N\rangle}$, where $\langle N\rangle$ is the expected
number of events. Since no events are detected in the pulsar timing
data, the upper limit on the expected number, $\langle N_*\rangle$, is
set so that $e^{-\langle N_*\rangle} = 0.05$. Hence, $\langle N\rangle
\le \langle N_*\rangle = 3$. If the actual expected number were
greater than $3$, then at least one source would have been detected
with 95\% probability.

Provided that the expected number of events that occur within the
resolution bandwidth is less than one, the expected number of
detectable events is given by
\begin{equation}
\langle N \rangle = \int\frac{d^2R}{d\lg(1+z) d\lg(M_c)}\left(\frac{df}{dt}\right)^{-1}P_d(M_c,z,f) d\lg(1+z) d\lg(M_c)df,
\label{exp_num}
\end{equation}
where $P_d(M_c,z,f)$ is the probability of detecting a SMBH binary
with chirp mass $M_c$ at a redshift of $z$ with observable frequency
$f$. $P_d$ takes into account non-GW noise sources that can reduce the
GW detection efficiency. Following the same argument used in the
previous section to obtain an upper bound on the differential rate
using an integral constraint, we find that:
\begin{equation}
\frac{d^2R}{d\lg(1+z)d\lg(M_c)}< \frac{15}{\int \left(\frac{df}{dt}\right)^{-1}P_d(M_c,z,f) df}.
\label{const_pn}
\end{equation}
As with the stochastic constraint, it is assumed that
$\Delta\lg(M_c)=1$ and $\Delta \lg(1+z) = 0.2$. This constraint should
be appropriately rescaled if the integrand in equation \ref{exp_num}
varies over shorter intervals. 

The above constraint requires a knowledge of $P_d$, the probability of
detecting a SMBH binary system using pulsar timing data. The $P_d$
is calculated using a method similar to that in \citet{yhj+10}.
For our analysis we use a Monte-Carlo simulation together with the
data analysis pipeline \citep{hjl+09} used to search for GWs in real and
simulated pulsar timing data.  We use a Neyman-Pearson decision
technique together with a Lomb periodogram to determine the
probability of detection. For a set of $N_p$ pulsars, the power
spectra of the timing residuals are calculated and added
together. This summed power spectrum is used as the detector.  We
determine noise levels and hence detection thresholds for each
frequency channel so that the false alarm rate for a detection is
0.001 across the entire power spectrum. The detection thresholds are
found by producing 100,000 fake data sets by shuffling the input
timing residuals, carrying out standard pulsar timing fits and forming
the summed power spectrum.

Once the thresholds are determined, the probability of detecting a GW
with a given strain amplitude is calculated as follows.  A GW strain
amplitude $A$ and frequency $f$ are chosen (this procedure is repeated
on a logarithmically spaced grid where $A$ ranges from
$10^{-16}$--$10^{-10}$ and $f$ ranges from 1/(30 years)--1/(2
weeks)). The GW polarization properties are chosen to be consistent
with GWs emitted from a binary system and simulated as described in
\citet{hjl+09}. The direction of the GW wave-vector is chosen from a
distribution that is uniform on the sky while its polarization is
drawn from a distribution of randomly oriented binary systems. The
induced timing residuals from this GW source are added to a shuffled
version of the original residuals and the summed periodogram is
calculated. This is repeated 1000 times and $P_d(A,f)$ is given by the
number of times that the GW was detected (i.e. produced power above
the threshold) divided by the total number of trials.

\section{Results and Discussion}

\begin{deluxetable}{lccc}
\tablecolumns{4}
\tablewidth{0pc}
\tablecaption{Upper limits on the amplitude of the stochastic GW
  background. A '--' indicates that there is no valid range for that scenario.
\label{range}}
\tablehead{
\colhead{Data set} & {$A_{\rm up}$} & \multicolumn{2}{c}{Valid redshift range} \\
\cline{3-4}
                   &               & {($10^9~\Msolar$)}& {($10^{10}~\Msolar$)}
}
\startdata
PPTA data \citep{jhv+06}&$1.1\times10^{-14}$& 0.01--180.08& 1.78--4.02 \\
20 PSRs-500\,ns-10\,yr & $1.1\times10^{-15}$& 0.03--602.60& -- \\ 
20 PSRs-100\,ns-5\,yr  & $9.9\times10^{-16}$& 0.07--294.97& -- \\ 
20 PSRs-100\,ns-10\,yr & $2.2\times10^{-16}$& 0.14--115.98& -- \\

20 PSRs-10\,ns-10\,yr  & $2.0\times10^{-17}$&      --     & -- \\
100 PSRs-100\,ns-5\,yr & $5.7\times10^{-16}$& 0.14--121.46& -- \\ 
100 PSRs-100\,ns-10\,yr& $1.3\times10^{-16}$& 0.27--45.30 & -- \\ 
100 PSRs-10\,ns-10\,yr & $8.8\times10^{-18}$&      --     & -- \\
\enddata 
\end{deluxetable}

The expressions in the previous sections can be used to provide
constraints on the rate of coalescence with any measurement of
$A_{\rm up}$ (for the stochastic case) or $P_d$ (for the Poissonian case).
Here we discuss the implications of the value of $A_{\rm up}$ presented by
\cite{jhv+06}, use the same data set to determine $P_d$ and simulate
data sets predicting possible future timing residuals.  For these
future data sets we simulate (1) a realistic goal for
existing pulsar timing array experiments (20 pulsars,
timed with an rms timing residual of 500\,ns over 10 years), (2) the
goal of the PPTA project (20 pulsars, timed with
an rms of 100\,ns over 5 years), and a more challenging goal of
20 pulsars, timed with an rms of 100\,ns over 10 years.  The rms
timing residuals that will be achieved with future telescopes, such as
the SKA, is not easy to determine.  It may be possible to time a few
pulsars with exquisite precision (with rms timing residuals of 10\,ns)
but other unmodeled noise processes may make this difficult or impossible. We
therefore also simulate the following possible future data sets: (4)
100 pulsars timed at 100\,ns over five years, (5) the same for ten
years, (6) 20 pulsars timed at 10\,ns for 10 years and (7) the same
for 100 pulsars.

\begin{figure*}
\epsscale{0.9}
\plotone{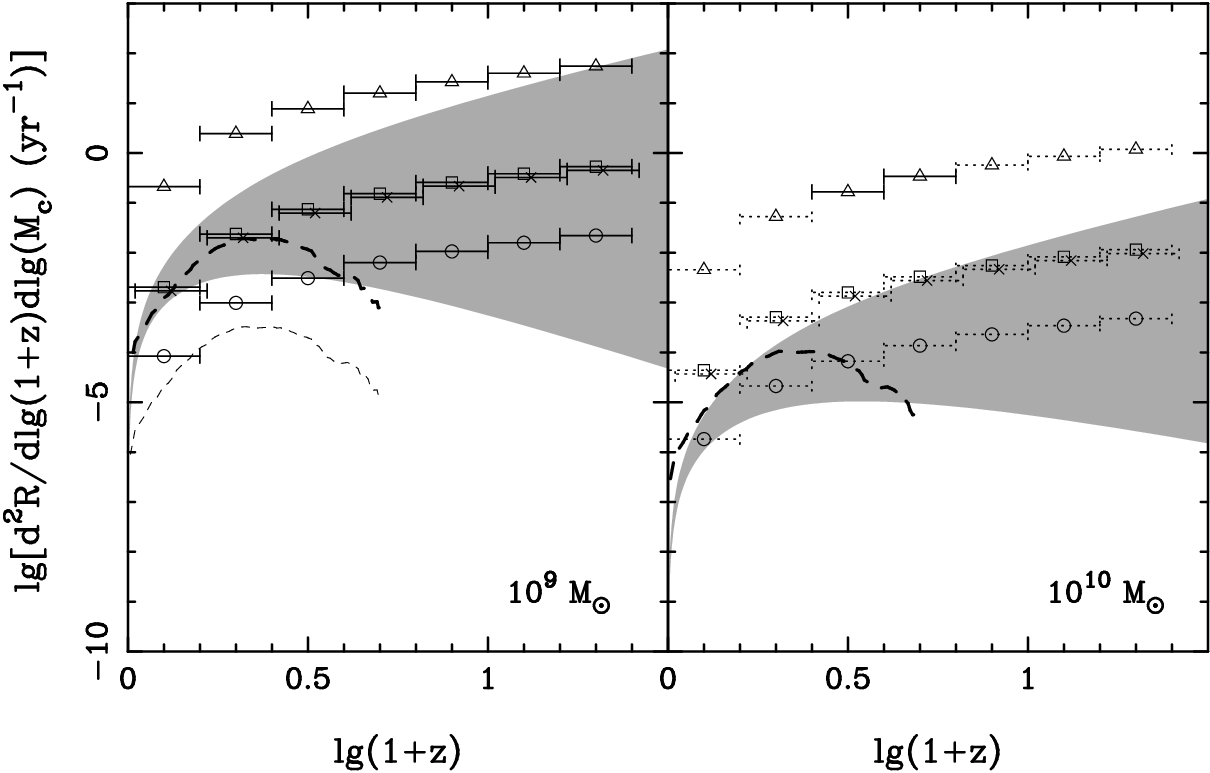}
\caption{Upper limits on the SMBH binary merger rate determined by the
stochastic constraint discussed in \S2.1 for different data sets: real
data published by \citet{jhv+06} (open triangle), simulated data for
20 PSRs-500\,ns-10\,yr (open square), 20 PSRs-100\,ns-5\,yr (cross),
and 20 PSRs-100\,ns-10\,yr (open circle).  The error bar is plotted as
a dotted line when the constraint is invalid. The filled gray area
represents the expected region for the coalescence rate using the
framework of \citet{jb03} together with the data from the SDSS
\citep{wlh09} with an evolution index  $-1 < \gamma < 3$. The dashed
lines between $0<\lg(1+z)<0.7$ indicate the maximum (thick) and
minimum (thin) predicted rates from \citet{svc08,svv09}.
  \label{rate_st1}}
\end{figure*}

\begin{figure*}
\epsscale{0.9}
\plotone{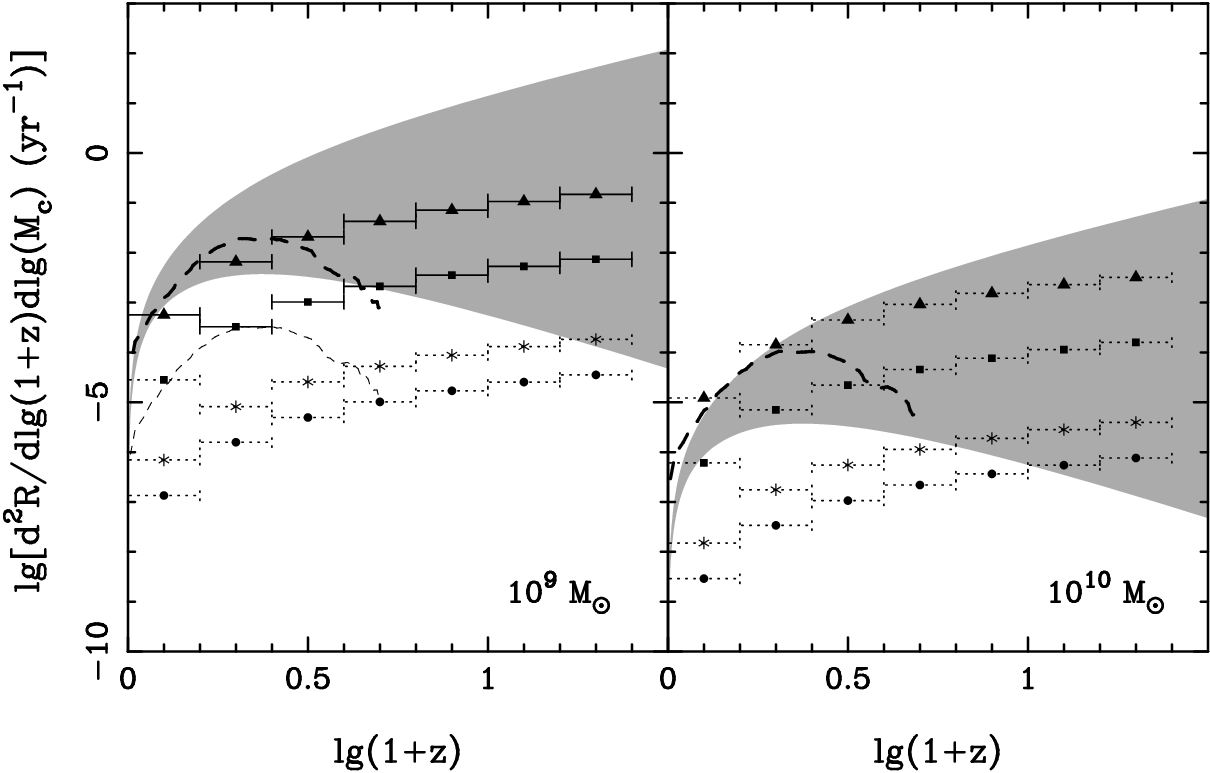}
\caption{Same as Figure~\ref{rate_st1}, but for simulated data sets: 
  100 PSRs-100\,ns-5\,yr (filled triangle), 100 PSRs-100\,ns-10\,yr
  (filled square), 20 PSRs-10\,ns-10\,yr (star), and 100 PSRs-10\,ns-10\,yr
  (filled circle).
\label{rate_st2}}
\end{figure*}

Figures ~\ref{rate_st1} and \ref{rate_st2} plot the stochastic
constraints given by equation~\ref{const_st} for the different
observing scenarios. The horizontal error bars indicate the region of
$\lg(1+z)$ over which the constraint is placed. The solid error bars
indicate that the stochastic constraint is valid, while the dotted
error bars indicate that the stochastic validity condition is violated
over the whole range. Table~\ref{range} gives the valid redshift range
for each data set and chirp mass. As discussed above, it was assumed
that $\Delta\lg(1+z)=0.2$ and $\Delta\lg(M_c)=1$ in order to calculate
the constraints shown. The constraints should be rescaled if other
values are assumed.

\begin{figure*}
\epsscale{0.9}
\plotone{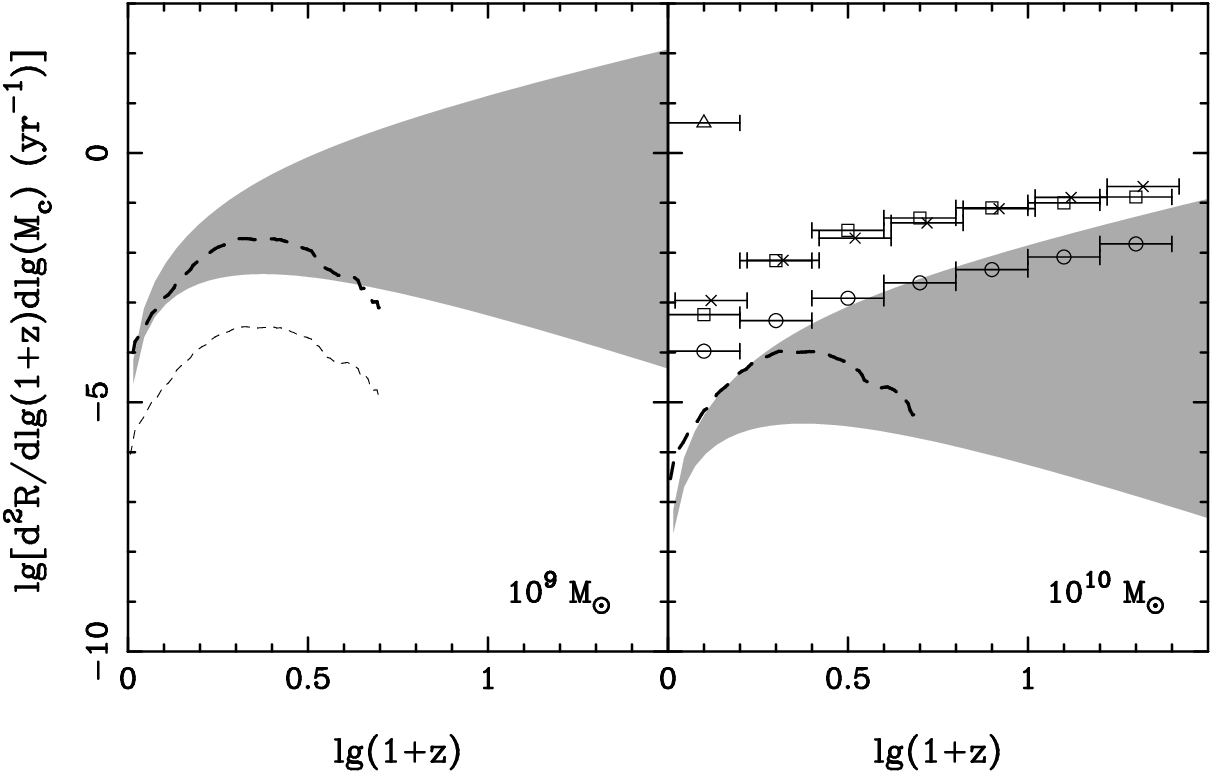}
\caption{Upper limits on the SMBH merger rate using the Poissonian
  constraint discussed in \S2.2 for different data sets: real data from \citet{jhv+06}
  (open triangle), 20 PSRs-500\,ns-10\,yr (open square), 20
  PSRs-100\,ns-5\,yr (cross), and 20 PSRs-100\,ns-10\,yr (open
  circle).  The filled gray area represents the expected region for
  the coalescence rate using the framework of \citet{jb03} together
  with the data from the SDSS \citep{wlh09} with an evolution index,
  $-1 < \gamma < 3$. The dashed lines between $0<\lg(1+z)<0.7$
  indicate the maximum (thick) and minimum (thin) predicted rates from
  \citet{svc08,svv09}. Note that no upper limits on the coalescence
  rate are obtainable for SMBH
  of $M_c=10^9~\Msolar$ at these sensitivities.
  \label{rate_pn1}}
\end{figure*}

\begin{figure*}
\epsscale{0.9}
\plotone{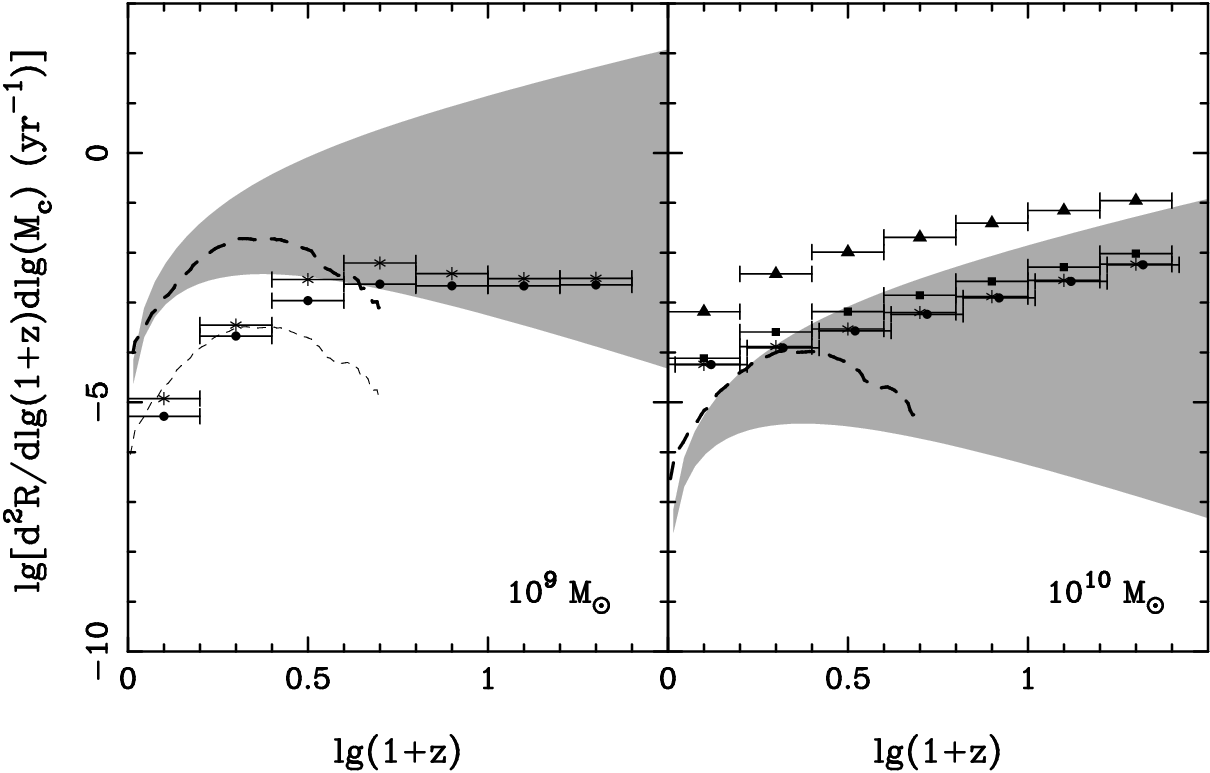}
\caption{Same as Figure~\ref{rate_pn1}, but for the following
  simulated data sets: 100 PSRs-100\,ns-5\,yr (filled triangle), 100 PSRs-100\,ns-10\,yr
  (filled square), 20 PSRs-10\,ns-10\,yr (star), and 100 PSRs-10\,ns-10\,yr
  (filled circle).
\label{rate_pn2}}
\end{figure*}

For redshifts where the stochastic technique is not valid, the
Poissonian constraint may be used.  Using $P_d(A,f)$,
Equation~\ref{amp} is used to calculate $P_d(M_c,z,f)$ and then the
right hand side of Equation~\ref{const_pn} is evaluated numerically to
determine the constraint on the differential coalescence rate. The
results are plotted versus redshift for different chirp masses in
Figures~\ref{rate_pn1} and \ref{rate_pn2}.  We note that the recently published
PPTA data set, that with 20 pulsars timed with an rms of 500\,ns over
10\,years, and that with 20 pulsars timed with an rms of 100 \,ns over
5\,years are not constraining for SMBH binaries with
$M_c=10^9~\Msolar$ and are therefore not plotted in the left-hand
panel of Figure~\ref{rate_pn1}.

In order to understand how constraining the pulsar rate limits are,
plots of the expected SMBH binary coalescence rate are also shown in
the figures. Several authors have developed analytical and numerical
techniques to estimate the coalescence rate of SMBH binaries. Here, we
compare the measured rate constraints to the rates predicted by the
models of \citet{jb03} and \citet{svc08,svv09}. For the case of
\citet{jb03}, the differential rate of SMBH binary coalescence is
given by
\begin{equation}
\frac{d^2R}{d\lg(1+z) d\lg(M_c)}=\frac{4\pi c}{H_0}\frac{D(z)^2}{\lg(e)E(z)}
\frac{\Phi(\lg (M_c))}{n}\Re(z),
\label{ratezm}
\end{equation}
where $\Re(z)$ is the total merger rate of SMBH binaries per unit comoving volume,
$\Phi(\lg(M_c))$ is the chirp mass distribution of merging SMBH binaries and $n$ is
the number density of SMBH binaries given by $n=\int \Phi(\lg(M_c))d\lg (M_c)$. It is
assumed that the merger rate of SMBH binaries is given by a fraction,
$\epsilon$, of the galaxy merger rate and that the rate evolves as a
power of $(1+z)$. Hence, one can write $\Re(z) = \epsilon \Re_g(0)(1
+ z)^\gamma$ where $\Re_g(0)$ is the local merger rate of galaxy pairs 
and $\gamma$ is the evolution index which is thought to be
within the range $-1 < \gamma < 3$
\citep[e.g.,][]{ppc+02,lkw+04,kss+07,lpk+08}. \citet{wlh09} determined
$\Re_g(0)$ and $\Phi(\lg(M_c))$ for luminous galaxies by analyzing data
from the Sloan Digital Sky Survey \citep[SDSS;][]{yaa+00}. They found
that $\Re_g(0) = (1.0\pm0.4)\times 10^{-5}~{\rm ~Mpc^{-3}~Gyr^{-1}}$ and 
\begin{equation}
\lg [\Phi(\lg (M_c))]=(21.7\pm4.2)-(3.0\pm0.5)\lg \left(M_c/\Msolar\right).
\label{chmass}
\end{equation}
Note that \citet{wlh09} showed that the rate implied by equation
\ref{ratezm} together with the above estimates for $\Re_g(0)$ and
$\Phi(\lg (M_c))$ yield an expected characteristic strain spectrum
consistent with other published estimates of $h_c$
\citep{wl03,shm+04,eins04,svc08}.

For the \citet{svc08,svv09} models, the rate 
may be estimated from the following expression: 
\begin{equation}
\frac{d^2R}{d\lg(1+z) d\lg(M_c)}= \frac{d\dot{N}_m}{d\lg(M_c)}
\frac{1}{\dot{N}}\frac{d\dot{N}}{dz}\frac{(1+z)}{\lg(e)}
\label{rate_sesana}
\end{equation}
where $d\dot{N}_m/d\lg(M_c)$ is the mass function of coalescing SMBHs in
the notation of \citet{svv09}, and $d \dot{N}/dz$ is the SMBH binary
coalescence rate per unit redshift in the notation of
\citet{svc08}. The constant $\dot{N}$ is given by
\begin{equation}
\dot{N} = \int_0^4 \frac{d\dot{N}}{dz} dz.
\end{equation}
The limits of the integral are set by the data presented in figure 12
of \citet{svc08}. The factor of $(1+z)/\lg(e)$ is used to convert the
differential $dz$ into $d\lg(1+z)$. From figure 1 of \citet{svv09}, we
can estimate the maximum and minimum predicted values of the mass
function $d\dot{N}_m/d\lg(M_c)$ over the four models presented
therein. We find that, for $M_c=10^9 \Msolar$, the mass function lies
between $10^{-4} ~\mbox{yr}^{-1}$ and $6\times 10^{-3}
~\mbox{yr}^{-1}$. These correspond to the ``Tr-SA'' and the ``La-SA''
models, respectively, as discussed in \citet{svv09}. For the case of
$M_c=10^{10} \Msolar$, the predicted range lies between $0$ an $3
\times 10^{-5} ~\mbox{yr}^{-1}$. These values also correspond to the 
``Tr-SA'' and ``La-SA'' models, respectively. For the SMBH binary coalescence rate
per unit redshift, $d\dot{N}/dz$, we used the data presented in
Figure 12 of \citet{svc08}. The three possible models shown are all
approximately within a factor of two of each other. For definiteness,
we chose the prediction based on the BVRhf model. In this case,
$\dot{N}~\approx 0.05 ~\mbox{yr}^{-1}$.

Figures \ref{rate_st1} and \ref{rate_st2} plot the pulsar timing
stochastic constraint together with the expected rates for the
theoretical models considered above as a function of redshift for
different chirp masses. The Poissonian constraint is plotted together
with the expected rates in Figures \ref{rate_pn1} and
\ref{rate_pn2}. Since there are few to no close SMBH binary systems
detected near $z=0$, it can be assumed that $\epsilon \approx 1$. The
only free parameter remaining in the \citet{jb03} model is the
evolution index which determines the SMBH binary merger rate as a
function of redshift. The grey regions in Figures
\ref{rate_st1}--\ref{rate_pn2} give the range of expected merger rates
for $-1 < \gamma < 3$.  The maximum and minimum rates found using the
\citet{svc08,svv09} models are shown as thick dashed and thin
dashed lines, respectively. Note, since the minimium predicted rate for
$M_c=10^{10} \Msolar$ is not presented by the Sesana et al. models, this curve is
not shown. Overall, the upper bounds obtained by pulsar timing data do
not contrain the parameters of the SMBH binary merger models discussed in
this paper beyond their currently accepted ranges. For the PPTA goal (20 pulsars timed at
100\,ns rms accuracy for five years), the results imply that either a
detection will be made or $\gamma<1.7$ at redshift $z<3$. In order to
place constraints that will limit the models of \citet{svc08,svv09} as
well as the \citet{jb03} model with $\gamma < -1$, one must either
time 100 pulsars with 100\,ns rms timing precision or 20 pulsars at
the 10\,ns level, both of which should be possible with the proposed
Square Kilometre Array project.

The Poissonian constraint is only useful for constraining the
properties of the most massive SMBH binaries since these systems are
rarer than their less massive counterparts and emit a stronger GW
signal. Figure \ref{rate_pn1} shows that the ideal PPTA extended to 10
years of observations will just be able to place useful limits for the
most massive systems if $\gamma$ were in the larger end of its
possible range. It will be more interesting when we can time pulsars
to the 10\,ns level. Here, the Poissonian constraint will be well
below that expected for $\sim$$10^9~\Msolar$ SMBH binaries from both
the \citet{svc08,svv09} and \citet{wlh09} models.  Hence, with 20
pulsars timed with 10\,ns rms timing accuracy for 10 years, we have a
very good chance of detecting an individual source or will place very
stringent constraints on models of SMBH binary formation and
evolution. These conclusions are consistent with the recent work of
\citet{svv09}.

\section{Summary}
\label{discuss}

We have shown that pulsar timing observations may be used to place
constraints on the rate of coalescence of binary supermassive black
holes distributed throughout the Universe. Two types of constraints
were considered: a stochastic constraint and a Poissonian
constraint. The stochastic constraint, which is based on a detection
algorithm for the stochastic GW background, gives lower rates but it
is only valid when the expected amplitude for a individual source is
much less than the minimum detectable amplitude. When this is not the
case, the Poissonian constraint must be used. This constraint is based
on a continuous-wave detection algorithm and it assumes that the
number of coalescence events are distributed according to a Poisson
distribution. In both cases, it is assumed that the differential rate
of coalescence varies sufficiently slowly over a range of chirp masses
and redshifts. The precise numerical value of the constraint depends
on the size of the interval over which the rate is assumed to be
nearly constant.

The implied rate constraint obtained from recently published data
together with rate constraints expected from future possible observing
scenarios were compared to theoretical rates calculated from different
models. It was shown that 20 pulsars timed with an accuracy of
100\,ns, the goal of the PPTA project, will place stringent
constraints on the semi-empirical models based on the work of
\citet{jb03} and \citet{wlh09}. The upper end of the range of
backgrounds produced by SMBH population synthesis models
\citep{svc08,svv09} would be detectable by the PPTA goal sensitivity,
but higher sensitivity will be needed to further constrain these models
if the GW background is not detected. It was also shown that if future
observations can time a pulsar with 10\,ns accuracy, a direct
detection of one or more individual sources is highly likely.

\acknowledgments

We thank Professor Han J. L. for his support and valuable comments. FJ
was supported by the National Science Foundation (CAREER grant no. AST
0545837). GH is the recipient of an Australian Research Council QEII
Fellowship (project \#DP0878388). ZW is supported by the National
Natural Science Foundation of China (projects \#10521001 and
\#10833003). FJ also thanks these projects for local support during
his visit at the NAOC for two months in 2006. The authors also wish to
thank the anonymous referee for all her/his useful suggestions and
comments during the review process. The Parkes telescope is part of
the Australia Telescope which is funded by the Commonwealth Government
for operation as a National Facility managed by CSIRO.



\end{document}